\documentclass{PoS}
\usepackage{graphicx,epstopdf,amsmath,amsfonts,amssymb,color}

\newcommand{\eV}{\ensuremath{\mathrm{eV}}}
\newcommand{\keV}{\ensuremath{\mathrm{keV}}}

\newcommand{\GeV}{\ensuremath{\mathrm{GeV}}}

\renewcommand{\day}{\ensuremath {\mathrm{day}}}

\title{Direct detection prospects of dark vectors with xenon-based
dark matter  experiments}

\ShortTitle{Dark Photon Dark Matter}

\author{Haipeng An\\
Walter Burke Institute for Theoretical Physics,
California Institute of Technology, Pasadena, CA 91125, USA\\
E-mail: \email{anhp@caltech.edu}
}

\author{Kaixuan Ni\\
Department of Physics, University of California San Diego, CA 92093, USA\\
E-mail: \email{nikx@physics.ucsd.edu}
}

\author{Maxim Pospelov\\
Department of Physics and Astronomy, University of Victoria, 
Victoria, BC V8P 5C2, Canada\\
Perimeter Institute for Theoretical Physics, Waterloo, ON N2J 2W9, 
Canada\\
E-mail: \email{mpospelov@perimeterinstitute.ca}
}

\author{Josef Pradler\\
Institute of High Energy Physics, Austrian Academy of
  Sciences,  1050 Vienna, Austria\\
E-mail: \email{josef.pradler@oeaw.ac.at}
}

\author{Adam Ritz\\
Department of Physics and Astronomy, University of Victoria, 
Victoria, BC V8P 5C2, Canada\\
E-mail: \email{aritz@uvic.ca}
}

\abstract{Dark matter experiments primarily search for the scattering
  of WIMPs on target nuclei of well shielded underground detectors.
  The results from liquid scintillator experiments furthermore provide
  precise probes of very light and very weakly coupled particles that
  may be absorbed by electrons. In these proceedings we summarize
  previously obtained constraints on long-lived dark matter vector
  particles $V$ (dark photons) in the $0.01-100$ keV mass range. In
  addition, we provide a first projected sensitivity reach for the
  upcoming XENON1T dark matter search to detect dark photons.}

\FullConference{The European Physical Society Conference on High Energy Physics\\
		22--29 July 2015\\
		Vienna, Austria}

\begin{document}

\section{Introduction}

The particle nature of Dark Matter (DM) is poorly understood and the
range of theoretical possibilities remains wide open. There are high
expectations that new physics exists at or near the electroweak scale,
for which a weakly interacting massive particle (WIMP) becomes a
viable DM option. Models of this type typically predict a significant
scattering rate for WIMPs in the galactic halo on nuclei, when up to
$100~{\rm keV}$ of WIMP kinetic energy can be transferred to
atoms. Such direct detection searches present a rapidly growing field
and particularly significant gains in sensitivity are to be expected
with upcoming ton-scale experiments~\cite{Aprile:2012zx,Akerib:2015cja}.

However, WIMPs are not the only possibility. Dark matter could be in
form of super-weakly interacting particles with masses well below the
electroweak scale. A prominent example of this type is the QCD
axion. Another example is that of a massive vector particle that
kinetically mixes with the Standard Model (SM) hypercharge field
strength, often referred to as ``dark photon''. Such forms of DM are
harder to detect directly, as the couplings to the SM are usually
smaller than those of WIMPs by many orders of magnitude.

The phenomenology of $\keV$-mass vector particles was first considered
in~\cite{Pospelov:2008jk, Redondo:2008ec}, where
in~\cite{Pospelov:2008jk} the sensitivity of liquid xenon experiments
to $\keV$-mass vector particles with couplings of $O(10^{-10})$ and
below was pointed out. Recently, this has been explored in greater
detail in~\cite{Arisaka:2012pb,An:2014twa} and here we summarize the
findings of~\cite{An:2014twa} which derive the leading limits to date
on dark photon dark matter in the mass window of 10~eV to several
hundreds of keV. In addition, we also provide a first sensitivity
projection for the future XENON1T experiment.

\section{Dark photon dark matter}

A technically natural extension of the SM is a new Abelian $U(1)'$
massive vector field that is coupled to SM hypercharge $U(1)$ through
kinetic mixing~\cite{Okun:1982xi,Holdom:1985ag}. For phenomena with involved
energies well below the electroweak scale, the mixing with the photon
is the most important,
\begin{align}
  \label{eq:L}
  \mathcal{L} = -\frac{1}{4} F_{\mu\nu}^2-\frac{1}{4} V_{\mu\nu}^2 -
  \frac{\kappa}{2} F_{\mu\nu}V^{\mu\nu} + \frac{m_V^2}{2} V_{\mu}V^{\mu}
  + e J_{\mathrm{em}}^{\mu} A_{\mu}.
\end{align}
The effective parameter $\kappa$ controls the coupling between the
dark photon ($V$) and photon ($A$) with respective field strengths
$V_{\mu\nu}$ and $F_{\mu\nu}$; $ J_{\mathrm{em}}^{\mu} $ is the
electromagnetic current and $m_V$ is the dark photon mass. In the
following, $m_V$ is assumed to added by hand (St\"uckelberg mass),
rather than being induced via a Higgs mechanism.
The search for dark photons has become a significant effort, and it
defines one of the prime targets of intensity frontier experiments,
see, \textit{e.g.}, \cite{Essig:2013lka} and references therein.

The cosmological abundance of $V$ with $m_V < 2 m_e$ receives various
contributions in the early Universe, such as production through
scattering or annihilation, $\gamma e^{\pm}\to V e^{\pm}$ and
$e^+e^-\to V\gamma$. If dark photons are to be dark matter, it turns
out that the thermal production is not sufficient to generate the
correct relic abundance in the mass window of interest.  However, dark
photon dark matter remains possible when the relic density receives
contributions from inflationary perturbations. Even in absence of any
initial misalignment, the gravitational production of $V$ can account
for the observed dark matter density in longitudinal
modes~\cite{Graham:2015rva},
\begin{align}
 \label{eq:omg-inflation}
  \Omega_V  \sim 0.3 \sqrt{\frac{m_V}{1\,\keV}} 
  \left( \frac{H_{\rm inf}}{10^{12}\,\GeV} \right) .
\end{align}
For keV vector particles, the relic density requirement then points to
an inflationary Hubble scale, $H_{\rm inf}$, in the $10^{12}\,\GeV$
ballpark. In the following we assume that dark matter is made from
dark photons and the galactic distribution is smooth and neglect any
effects from substructure.

\section{Absorption in liquid xenon experiments}
\label{sec:dm-absorpt-sign}

Galactic dark photon dark matter can induce ionization of xenon atoms
in the encounter with the detector when their mass exceeds the binding
energy of electrons of the outermost atomic shell,
$m_V \simeq E_V \gtrsim 12\,\eV$,
\begin{equation}
{\rm Xe~I} + V \to {\rm Xe~II} +e^- .
\label{ion} 
\end{equation}
It is of course possible that multiple electrons are produced in the
absorption. The electron multiplicity is of great importance when
seeking the discrimination of a potential dark photon signal from
other electromagnetic backgrounds. For the purpose of setting
conservative constraints, however, we are allowed to neglect such
complications~\cite{An:2013yua,An:2014twa} but note the improvement
potential for future studies.

At non-relativistic relative velocities the distinction between
longitudinal and transverse modes disappears and that the polarization
state of $V$ is inconsequential.  It also turns out that matter
effects in liquid xenon are of little importance when considering the
absorption (\ref{ion}); the vacuum mixing angle $\kappa$ times the
electric charge $e$ controls the coupling to electrons and hence the
rate of ionization. Restricting our attention to electric dipole (E1)
transitions provides a reasonably good approximation to the absorption
cross section~\cite{Pospelov:2008jk},
\begin{equation}
\sigma_V(E_V=m_V)v_{V} \simeq \kappa^2\sigma_\gamma(\omega=m_V)c,
\label{relation}
\end{equation}
where $v_{V}$ is the velocity of the incoming DM particle,
$\sigma_{\gamma}$ is the ordinary photon ionization cross section, and
$\omega$ is the photon energy. Improvements to the estimate
(\ref{relation}) are possible through atomic theory calculations; for
the case of axion-like DM, these have already been
performed~\cite{Dzuba:2010cw}.  The expression (\ref{relation}) is
nearly independent of the dark photon velocity, with the consequence
that the possibly intricate DM velocity distribution is of almost no
importance; this is in stark contrast to the case of elastic
scattering of electroweak-mass dark matter on nuclei.
In our numerical calculations, we employ the optical theorem which
includes matter effects (for a derivation see the original works
\cite{An:2013yua,An:2014twa}.) The total absorption rate of
non-relativistic dark photons with $\keV$-masses in the lab-frame of
the detector is then given by~\cite{An:2013yua,An:2014twa},
\begin{equation}
\Gamma \simeq \kappa^2 \omega \times  {\rm Im }~ n_{\mathrm{refr}}^2 
 = \kappa^2 \sigma_{\gamma}\times \left(\frac{N_{\rm at}}{V} \right),
\end{equation}
Here, $n_{\mathrm{refr}} $ is the (complex) refractive index of liquid
xenon which we obtain from tabulated atomic scattering
data~\cite{Henke:1993eda}; $N_{\rm at}$ is the number of target nuclei
in the fiducial detector volume $V$. 
\begin{figure}[tb]
\centering
\includegraphics[width=0.6\textwidth]{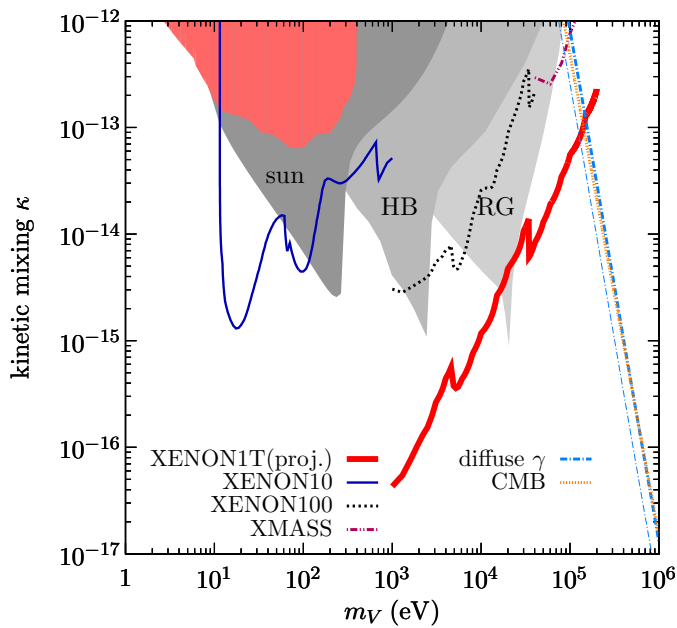}%
\caption{\footnotesize A summary of constraints on the dark photon
  kinetic mixing parameter $\kappa$ as a function of vector mass
  $m_V$. The regions above the thin lines are excluded for dark photon
  dark matter.  Shaded regions are astrophysical limits that are
  independent of the dark photon relic density;
  see~\cite{An:2013yua,An:2014twa} for details. The thick red line is
  a projected sensitivity curve for the upcoming XENON1T experiment
  (see main text for details.)  }
\label{fig:summary}
\end{figure}

The ensuing limits using the results from the XENON10 and XENON100
experiments are shown in Fig.~\ref{fig:summary}.  The XENON10 limit is
obtained from the number of detected electrons in a relatively small
data set with 15~kg-days exposure; similar ionization-only analyses
were previously performed to constrain WIMP-electron scattering
\cite{Essig:2011nj,Essig:2012yx}. The XENON100 limit is obtained from
the scintillation signal S1 with an exposure of 224.6 live days and an
active target mass of 34~kg liquid
xenon~\cite{Aprile:2014eoa}. Finally, the XMASS limit is taken
from~\cite{Abe:2014zcd}. In addition to the direct laboratory limits,
dark photon dark matter with $m_V\gtrsim 100\,\keV$ is constrained by
observations of the diffuse $\gamma$-ray flux from $V\to 3\gamma$
decays as well as by energy injection during recombination (CMB). All
these limits, together with the astrophysical constraints that are
independent of the dark photon relic density are described in detail
in~\cite{An:2014twa}.

We expect that the sensitivity to the kinetic mixing below 1~keV will
be further improved by the upcoming ionization-only analyses from
XENON100. The XENON1T project, currently under final integration with
ultra-low radioactive materials and self-shielding with volume
fiducialization, is expected to reduce the electron recoil background
by two orders of magnitude compared to that of XENON100.  The electron
recoil background of XENON1T in the one-ton fiducial volume is mostly
from solar neutrinos, $^{85}$Kr, $^{222}$Rn and material
radioactivities below 25~keV and from $^{136}$Xe above 25~keV and up
to 200~keV~\cite{Aprile:APS2015}. The background spectrum shape is
almost flat and feature-less, in contrast to a energy peak at position
$m_V$ from dark photon absorption. Following the calculation for the
XENON100 limit~\cite{An:2014twa}, we derive the projected sensitivity
of XENON1T for a dark photon mass above 1~keV, shown in
Fig.~\ref{fig:summary}.  The simulated background of the XENON1T
detector used in the derivation is from Ref.~\cite{Aprile:APS2015},
and the energy resolution is assumed to be the same as that in
XENON100 as shown in Ref.~\cite{Kish:2011eqa}. In this analysis, we
assume the detection efficiency to be 100\% down to 1~keV. The
sensitivity for dark photon masses below 1~keV depends on the energy
threshold of scintillation light and the rate of ionization-only
events, which will be available once the experiment starts operating.

\section{Conclusions}
\label{sec:summary}

The model of light, kinetically mixed hidden vector particles with
St\"uckelberg mass is particularly simple and owing to its
UV-completeness, well-motivated.  These proceedings summarize the
results on dark photon dark matter that some of us have obtained
in~\cite{An:2014twa}.  In addition, we provide an update that shows
the potential sensitivity gain with ton-scale direct detection
experiments focusing on the example of XENON1T.  Making reasonable
assumptions on the expected background rates, more than an order of
magnitude improvement on the limit of the kinetic mixing parameter
$\kappa$ is possible, constraining electron-dark photon effective
couplings that correspond to a dark fine-structure constant as small
as $\alpha_{\rm dark} = \kappa^2 \alpha \sim 10^{-35}$.

\end{document}